\documentclass[reprint,amsmath,amssymb,aps,superscriptaddress,pra]{revtex4-2}

\usepackage{subfigure}
\usepackage{comment}
\usepackage{bm}
\usepackage{amsmath,amssymb}
\usepackage{amsthm}
\usepackage{mathrsfs}
\usepackage{braket}
\usepackage[utf8]{inputenc}

\usepackage{graphicx}

\usepackage{xcolor,graphicx} 

\usepackage{multirow}

\newcommand\D{\!\operatorname{d}\!}
\newcommand{\db}{\mathbf{d}}
\providecommand{\x}{\mathbf{x}}
\providecommand{\kk}{\mathbf{k}}
\newcommand{\rr}{\mathbf{r}}
\providecommand{\q}{\mathbf{q}}
\providecommand{\p}{\mathbf{p}}
\providecommand{\y}{\mathbf{y}}
\providecommand{\z}{\mathbf{z}}
\newcommand{\ab}{\mathbf{a}}
\newcommand{\bb}{\mathbf{b}}
\newcommand{\cb}{\mathbf{c}}
\newcommand{\nnu}{{\bm \nu}}
\newcommand{\mmu}{{\bm \mu}}
\providecommand{\ba}{\mathbf{a}}
\newcommand{\comm}[2]{\left[#1,#2\right]}
\newcommand{\Tr}[1]{\operatorname{Tr}\left[#1\right]}
\newcommand{\erf}{\operatorname{erf}\!}

\definecolor{junglegreen}{rgb}{0.16, 0.67, 0.53}
\definecolor{brickred}{rgb}{0.8, 0.25, 0.33}

\usepackage[normalem]{ulem}

\usepackage{hyperref}
\hypersetup{
 colorlinks=true,
 citecolor=red,
 linkcolor=blue,
 urlcolor=blue,
 pdfpagemode=UseNone,
 pdfstartview=FitH}

\footnotetext{These authors contributed equally.}

\begin{document}

\title{On the effectiveness of the collapse in the Di\'osi-Penrose model}

\author{Laria Figurato$^\diamond$}
\email{laria.figurato@phd.units.it}
\affiliation{Department of Physics, University of Trieste, Strada Costiera 11, 34151 Trieste, Italy}
\affiliation{Istituto Nazionale di Fisica Nucleare, Trieste Section, Via Valerio 2, 34127 Trieste, Italy}

\author{Marco Dirindin$^\diamond$}
\affiliation{Department of Physics, University of Trieste, Strada Costiera 11, 34151 Trieste, Italy}

\author{José Luis Gaona-Reyes}
\affiliation{Department of Physics, University of Trieste, Strada Costiera 11, 34151 Trieste, Italy}
\affiliation{Istituto Nazionale di Fisica Nucleare, Trieste Section, Via Valerio 2, 34127 Trieste, Italy}

\author{Matteo Carlesso}
\affiliation{Department of Physics, University of Trieste, Strada Costiera 11, 34151 Trieste, Italy}
\affiliation{Istituto Nazionale di Fisica Nucleare, Trieste Section, Via Valerio 2, 34127 Trieste, Italy}

\author{Angelo Bassi}
\affiliation{Department of Physics, University of Trieste, Strada Costiera 11, 34151 Trieste, Italy}
\affiliation{Istituto Nazionale di Fisica Nucleare, Trieste Section, Via Valerio 2, 34127 Trieste, Italy}

\author{Sandro Donadi}
\email{s.donadi@qub.ac.uk}
\affiliation{Centre for Quantum Materials and Technologies,
School of Mathematics and Physics, Queens University, Belfast BT7 1NN, United Kingdom}
\affiliation{Istituto Nazionale di Fisica Nucleare, Trieste Section, Via Valerio 2, 34127 Trieste, Italy}

\date{\today}
\begin{abstract}
The possibility that gravity plays a role in the collapse of the quantum wave function has been considered in the literature, and it is of relevance not only because it would provide a solution to the measurement problem in quantum theory, but also because it would give a new and unexpected twist to the search for a unified theory of quantum and gravitational phenomena, possibly overcoming the current impasse. The Di\'osi-Penrose model is  the most popular incarnation of this idea. It predicts a progressive breakdown of quantum superpositions when the mass of the system increases; as such, it is susceptible to experimental verification. Current experiments set a lower bound $R_0\gtrsim 4$\,\AA\, for the free parameter of the model, excluding some versions of it. In this work we search for an upper bound, coming from the request that the collapse is effective enough to guarantee classicality at the macroscopic scale: we find out that not all macroscopic systems collapse effectively. If one relaxes this request, a reasonable (although to some degree arbitrary) bound is found to be: $R_0\lesssim 10^6$\,\AA. This will serve to better direct future experiments to further test the model.
\end{abstract}

\maketitle

{\it Introduction.} -- 
Quantum Mechanics (QM) and General Relativity (GR) are the two pillars of modern physics, the former governing the microscopic world and the latter the macroscopic one, with a significant overlap. Yet,
a unified description is still missing and also experimentally we are essentially ignorant about the behaviour of gravity in the quantum domain. 
For example we still do not know what the gravitational field generated by a spatial superposition of matter is. 

The standard and by far most  studied approach to unification consists in quantizing gravity \cite{dewitt1967quantum,DIOSI1984199,hooft1993dimensional,Rovelli1998loop,green2012superstring,bell2016quantum}. But, in spite of many important results, a comprehensive theory of Quantum Gravity is still missing. More recently, an alternative  approach is slowly emerging, aiming at modifying the structure of QM in order to accommodate gravitational effects \cite{karolyhazy1966gravitation,ellis1984search,kafri2014classical,tilloy2016sourcing,bassi2017gravitational,oppenheim2023gravitationally,oppenheim2023postquantum}, which predicts the existence of extra (space-time) fluctuations. This approach is also motivated by the desire of finding a solution to the well-known quantum measurement problem \cite{bell2004speakable}.

One of the best examples of this line of research is provided by the Di\'osi-Penrose (DP) model \cite{diosi1987universal,diosi1989models,penrose1996,penrose2000,penrose2014gravitization} (see \cite{bahrami2014role,howl2019exploring,di2023linear,di2024experimental,D4CP02364A} for follow ups) and  belongs to a wider class of so-called collapse models \cite{ghirardi1986,ghirardi1990markov,bassi2003dynamical,bassi2013,carlesso2022present}. The DP model predicts the existence of a spontaneous collapse related to gravity.  As the mass of the quantum system increases and its associated gravitational field strengthens, the likelihood of spatial superposition of collapsing into well-localized states increases. As described below, the model is parametrized in terms of a length $R_0$, which smears the mass distribution of point particles. Equivalently, it can be seen as a high energy cutoff determining the strength of the collapse.

The different dynamics of the DP model with respect to QM allows testing it, placing experimental bounds on $R_0$. Specifically, these will be lower bounds, since the smaller the value of $R_0$, the higher the cutoff and the stronger the collapse affecting the wave function. Several experiments have been considered: matter-wave interferometry \cite{torovs2018bounds}, neutron star heating \cite{tilloy2019neutron}, gravitational waves detector \cite{helou2017lisa}, heat leakage in experiments with ultralow temperature cryostats \cite{vinante2021gravity} and spontaneous emission of photons from bulk materials \cite{donadi2021underground,arnquist2022search}. {Currently, the strongest bound is $R_0\gtrsim 4\times 10^{-10}$\,m \cite{arnquist2022search}, which has excluded one version of the model.} 

While experiments keep running,
it would help to constrain as much as possible the parameter space in order to better direct future tests.
Here we investigate possible upper bounds, as already considered for another collapse model, the Continuous Spontaneous Localization (CSL) model \cite{ghirardi1990continuous, torovs2018bounds}. The theoretical bound comes from the requirement that the collapse dynamics serves its main purpose of justifying why we do not see macroscopic quantum superpositions, thus solving the quantum measurement problem. Therefore, one can claim that --- to the least --- it should collapse quantum superpositions of any macroscopic object before we can potentially see them. Clearly there is a lot of ambiguity in this request, given that the macroscopic domain cannot be defined precisely. We can try to remove part of this ambiguity by sharpening the request as follows: a collapse model explains classicality of the macroscopic world if a delocalized state of the smallest object which can be directly seen by the human eye collapses within its perception time. If this happens, the quantum state of any larger object will collapse even faster and classicality is secured.

Our analysis will show that the DP model, strictly speaking, does not satisfy this requirement and therefore it does not guarantee classicality. However, we will also show that the collapse becomes rapidly effective as soon as the mass of the object increases.\\

\textit{The model. --} The master equation describing the time evolution for the statistical operator $\hat \rho(t)$ of a generic $N$-particle system under the DP dynamics is given by \begin{equation}\label{eq.master}
    \frac{\D}{\D t}\hat \rho(t)=-\frac i \hbar\comm{\hat H_N}{\hat \rho(t)}+\mathcal D[\hat \rho(t)],
\end{equation}
where $\hat H_N$ is the standard $N$-particle Hamiltonian  and 
\begin{equation}\label{me}
    \mathcal D[\hat \rho(t)]=-\frac{4\pi G}{\hbar}\int \D \rr\,\int\D\rr'\,\frac{1}{|\rr-\rr'|}\comm{\hat \mu(\rr')}{\comm{\hat \mu(\rr)}{\hat \rho(t)}},
\end{equation}
describes the DP non-unitary term. This term is the one  responsible for the collapse, {whose strength is determined by the Newtonian interaction (with $G$ the gravitational constant) and by the mass density operator $\hat \mu(\rr)$ of the system}. To avoid the standard divergences that occur when dealing with point-like masses, the mass density is smeared out over a width $R_0$, which is the only free parameter of the DP model. Specifically, by assuming for convenience that the mass density as well as the smearing are of a Gaussian form \cite{ghirardi1990continuous}, for a system of $N$ distinguishable particles of radius $R_i$, mass $m_i$ and of position operator $\hat \x_i$, one has 
\begin{equation}\label{def.hat.mu}
    \hat \mu(\rr)=\sum_{i=1}^N \frac{m_i}{(2\pi R_{\text{eff},i}^2)^{3/2}}e^{-\frac{(\rr-\hat \x_i)^2}{2R_{\text{eff},i}^2}},
\end{equation}
where $R_{\text{eff},i}=\sqrt{R_0^2+R_i^2}$ is the effective radius of the particle. In the limit of a point-like particle,  $R_{\text{eff},i}$ reduces to $R_0$. In the following, we will focus on the dynamics of the center-of-mass (CM) degrees of freedom alone, which can be obtained {by tracing Eq.~\eqref{eq.master} over the relative degrees of freedom}. When working with solid state materials, for values of $R_0$ which are not experimentally ruled out, one can safely neglect the contributions from electrons and focus on that of the nuclei because of the mass scaling of Eq.~\eqref{def.hat.mu}.

We assume that the system's CM is free ($\hat H_\text{\tiny CM}=\tfrac{\hat p^2}{2M}$, with $M=\sum_i m_i$). In Appendix A we show that for suitably short times and large superposition distances, one can neglect the Hamiltonian contribution to the dynamics ($\hat H_\text{\tiny CM}\to0$). Correspondingly, when represented in the position basis, the CM density matrix evolves according to
\begin{equation}\label{offdiagonal}
    \braket{\x|\hat \rho_\text{\tiny CM}(t)|\y}\simeq \braket{\x|\hat \rho_\text{\tiny CM}(0)|\y}\exp\left[-t/\tau(\x-\y)\right],
\end{equation}
with $\tau(\db)=\hbar/\Delta E(\db)$, where
\begin{equation} \label{DE_def}
    \Delta E(\db)=-8\pi G\int\D\rr\!\int\D\rr'\,\frac{\mu(\rr)\left[\mu(\rr'+\db)-\mu(\rr')\right]}{|\rr-\rr'|},
\end{equation}
quantifies, in the Newtonian limit, the difference between the space-time curvatures generated by two well localised configurations superimposed at a distance $d=|\db|$. Here,  $\mu(\rr)$ takes the form of Eq.~\eqref{def.hat.mu} with $\hat \x_i$ replaced by the coordinate $\x_i$. Merging this equation with the explicit form of $\mu(\rr)$, and assuming that the $N$ particles have the same mass $m$, we obtain
\begin{equation}\label{DeltaEij}
    \Delta E(\db)=8\pi Gm^2\sum_{i=1}^N\sum_{j=1}^Nf(\rr_{ij},R_0,\db),
\end{equation}
where 
\begin{equation}
 f(\rr_{ij},R_0,\db)=   \frac{\erf\left(\frac{r_{ij}}{2R_\text{eff}}\right)}{r_{ij}}-\frac{\erf\left(\frac{|\db-\rr_{ij}|}{2R_\text{eff}}\right)}{|\db-\rr_{ij}|},
\end{equation}
with $\rr_{ij}=\x_i-\x_j$ and $r_{ij}=|\rr_{ij}|$.\\ 

\textit{Application. --} We now fix the system of interest that will be used to quantify the theoretical upper bound on $R_0$. We consider a plate made of a single layer of graphene \cite{blake2007making}, with a side length of $L=25\,\mu$m corresponding to about the smallest space resolution of the eye \cite{jacobs2003comparative}. 
For the sake of simplicity, one can take the system as having a square shape, corresponding to a total {number of carbon atoms} $N=2\times 10^{10}$. We assume that the CM of the system is prepared in a spatial superposition of two wavepackets of width $\sigma$ at a superposition distance $d=|\db|$, with $\sigma\ll d$, where $d$ has been fixed to $4L=100\,\mu$m. 
This value is such that the two spatial configurations are distinguishable (namely, $d>L$ and $d>\sigma$), which is the requirement to have an actual cat state. As discussed in the Introduction, the goal is to compute the collapse time $\tau(\db)$ of this superposition and compare it with the resolution time of the human eye, $\tau_\text{obs}=0.01\,$s \cite{jacobs2003comparative}.   

We can  quantify $\tau(\db)$ by explicitly evaluating the sum in Eq.~\eqref{DeltaEij}. Given the large value of $N$, which includes $N^2$ terms, a direct summation is not feasible. Nevertheless, since $f(\rr_{ij},R_0,\db)$ depends only on the relative distance $\rr_{ij}$, we can rewrite the double sum in Eq.~\eqref{DeltaEij} as a single weighted sum. Namely,
\begin{equation}\label{doublesum}
\sum_{i=1}^N\sum_{j=1}^Nf(\rr_{ij},R_0,\db)=\sum_{\rr\in\mathcal D}\omega(\rr)f(\rr,R_0,\db),
\end{equation}
where $\omega(\rr)$ is the number of atoms at a distance $\rr$ within the
domain $\mathcal D$, which accounts for every possible distance among the atoms in the considered system. Correspondingly, the calculation strongly simplifies once considering that the atoms lie in a periodic lattice, and thus $\rr$ becomes a function of the primitive vectors $\ab_i$ and lattice index $n_i$.
For a two-dimensional lattice with primitive vectors $\ab_1$ and $\ab_2$ and respectively $N_1$ and $N_2$ sites along these directions (where $N_1N_2=N$), the domain is given by
\begin{equation}\label{domainNN}
    \mathcal D\!=\!\set{\!|n_1\ab_1+n_2\ab_2|\text{, where } n_i\in[-(N_i-1),(N_i-1)]\!}\!,
\end{equation}
and the weights read
\begin{equation}   \label{omegaNN}
    \omega(n_1,n_2)=(N_1-|n_1|)(N_2-|n_2|).
\end{equation}
Thus, the Eq.~\eqref{doublesum} becomes
\begin{equation}\label{numericsum}
    \sum_{n_1=-(N_1-1)}^{N_1-1}    \sum_{n_2=-(N_2-1)}^{N_2-1}\omega(n_1,n_2)f(n_1\ab_1+n_2\ab_2, R_0,\db),
\end{equation}
which involves the sum of $(2N_1-1)(2N_2-1)\sim 4N_1N_2=4N$ terms. Then, a speedup from a quadratic to a linear scaling in $N$ is obtained, which is  significant for $N\sim10^{10}$. To be quantitative, we estimated in Appendix D that the evaluation of the sum for $N=2\times 10^{10}$ takes around 5 minutes on a personal laptop by using Eq.~\eqref{numericsum}, while it would take around 13000 centuries by using the left-hand-side of Eq.~\eqref{doublesum}: see Appendix D, where we derive explicitly the expressions in Eq.~\eqref{domainNN} and Eq.~\eqref{omegaNN}, and extend this approach to a more general lattice.
The expression in Eq.~\eqref{numericsum}, and thus $\tau(\db)$, is computed numerically  for different values of $R_0$ and $N$, and by accounting for the actual geometry of the graphene lattice. We verified the numerical computation with analytical estimations that are derived in Appendix B.

\begin{figure}[t!] 
\centering
\includegraphics[width=\linewidth]{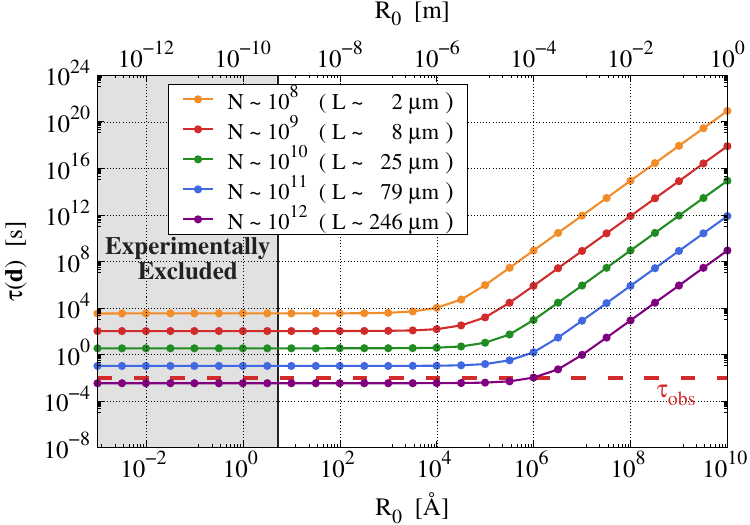} 
\caption{$\tau(\db)$ as a function of $R_0$ for different sizes (equivalently, number of atoms) of a graphene square plate. The plate is assumed to be in a superposition of two localized states at a distance $|\db|=4L$. The red dashed line indicates $\tau_\text{obs}$. Values of $R_0$ for which $\tau(\db)>\tau_\text{obs}$ do not guarantee classicality. We also report the values that are currently experimentally excluded (gray shaded region) \cite{arnquist2022search}.
} 
\label{exclusionplot}
\end{figure}

Figure~\ref{exclusionplot} shows the value of $\tau(\db)$ compared with $\tau_\text{obs}$. According to the analysis, the chosen graphene plate (corresponding to the green line) collapses on a timescale that is two orders of magnitude larger than $\tau_\text{obs}$, for any value of $R_0$. The model does not collapse the plate fast enough with respect to the resolution time of the human eye.

The analysis also shows that, to make the collapse time shorter than $\tau_\text{obs}$, one should take $\tau_\text{obs} \sim 3$\,s, which is a too long time for requiring a macroscopic object to localize. Therefore our result is robust against changes of $\tau_\text{obs}$.

\begin{figure}[t!] 
\centering
\includegraphics[width=\linewidth]{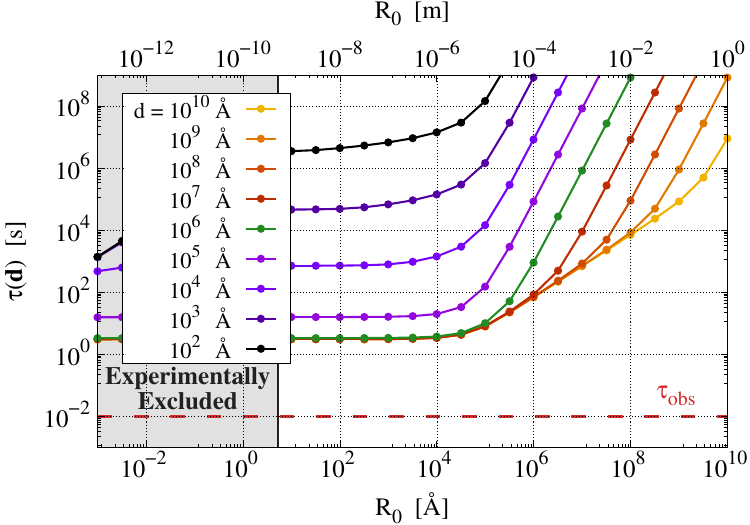} 
\caption{$\tau(\db)$ as a function of $R_0$ for different values of the superposition distance $d$, for a graphene square plate with $N=2\times10^{10}$ atoms (length $L=25 \mu$m). The value of $d=4L=10^6$\,\AA\, (green line) corresponds to that studied in Fig.~\ref{exclusionplot} (green line). The gray region is excluded experimentally. } 
\label{exclusionplot_vary_d}
\end{figure}

A natural question is what happens if the superposition distance $d=|\db|$ changes. Figure \ref{exclusionplot_vary_d} reports $\tau(\db)$ for different values of $d$, while keeping the size of the system fixed ($L=25\mu$m and $N=2\times 10^{10}$). With respect to the case studied in Fig.~\ref{exclusionplot} where $d=4L=100\,\mu$m (red line in Fig.~\ref{exclusionplot_vary_d}), we see that the collapse effect becomes stronger for larger values of $d$ but only for values of $R_0>L$. Conversely, for smaller values of $d$, the collapse effect loses strength for all values of $R_0$. In all cases, the collapse is not fast enough to occur before $\tau_\text{obs}$. Therefore our result is robust against changes in the delocalization distance.

The next relevant question is how the collapse time $\tau(\db)$ changes when modifying the system. We first checked what happens when increasing the size of the plate. Fig.~\ref{exclusionplot} shows that $\tau(\db)$ becomes smaller than $\tau_\text{obs}$ when the system length $L$ is about ten times larger than the resolution distance of the human eye ($L=246\,\mu m$, purple line in Fig.~\ref{exclusionplot}). In this case, we obtain an upper bound on $R_0$ at $10^{-4}\,\text{m}\,(=10^6$\,{\AA}).

\begin{figure}[t!] 
\centering
\includegraphics[width=\linewidth]{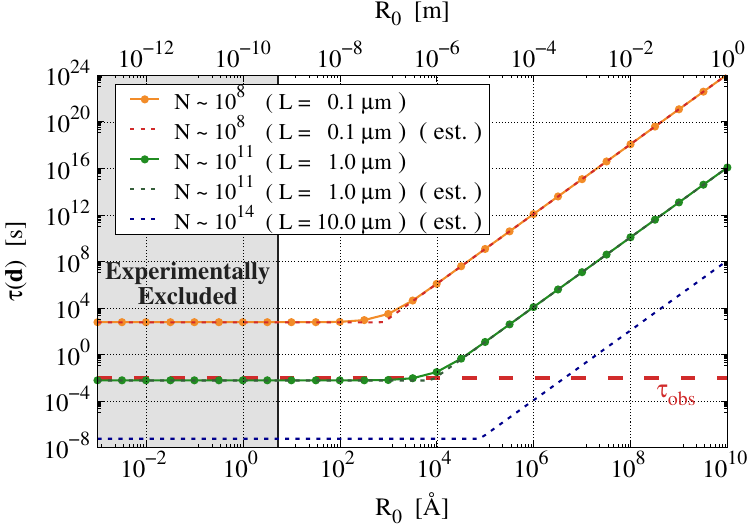} 
\caption{$\tau(\db)$ as a function of $R_0$ for different number of atoms $N$ (and thus length $L$) of a cubic graphene system in a superposition of two states at distance $|\db|=4L$. The dashed curve provides an estimate of $\tau(\db)$ (see SM).  The gray region is excluded experimentally.} 
\label{exclusionplot_3d}
\end{figure}

Next we study what happens when the system has a three dimensional structure. Specifically, we consider a cubic system made of stacked layers of graphene of length $L$ and require again that the DP dynamics collapses it on the time-scale $\tau_\text{obs}=0.01\,$s. The analysis is analogous to that performed in the two-dimensional case and the results are shown in Fig.~\ref{exclusionplot_3d}. 
In particular, we see that a system with  {$L\simeq1.0\,\mu$m} is sufficiently large to collapse within $\tau_\text{obs}$ for $R_0\lesssim10^3\,$\AA. We estimate that for an object of $L\simeq10\,\mu$m, such as that considered in Ref.~\cite{marshall2003towards}, the collapse takes place much faster (with {$\tau(\db)\simeq 6\times 10^{-8}\,$s} at the plateau) and up to a value of $R_0\lesssim5\times 10^6$\,\AA. Such an  estimation is provided by analytical calculations since the numerical approach is impractical for such a large number of atoms $N\sim10^{14}$ (see Appendix C)). We checked that the theoretical estimates match the numerical ones for smaller $N$.
The reason why a three-dimensional system collapses much more effectively compared to the case of a plate considered above is that, for the same length $l$, one has many more atoms involved, $\sim(L/a)^3$ vs $\sim(L/a)^2$, where $a$ is the lattice step. Moreover, when taking the same number of atoms, the atoms are much more densely disposed in a cube than in a plate, and this makes the collapse strength grow much faster due to  Newton's law in Eq.~\eqref{me}.\\

\textit{Modifications. --} Last,
we study how robust the previous analysis is with respect to physically-motivated modifications of the model. Specifically, in the standard DP model, one assumes the noise responsible for the collapse to be white, with a Dirac-delta as the two-time correlation function. Assuming that the collapse dynamics has a physical origin, a more realistic, colored noise should be considered. The latter will be characterised by a non-trivial two-time correlation function $f(t)$.
Details in the derivation of the colored collapse dynamics can be found in the literature \cite{diosi1997nonmarkovian,bassi2002Gaussiannoises,adler2007collapse}. Here we report the corresponding dissipator $\mathcal D[\hat \rho(t)]$ approximated to the second-order expansion in the noise, which reads
\begin{equation}\label{NMME}
\begin{aligned}
     \mathcal D[\hat \rho_\text{\tiny CM}(t)]=-\frac{8\pi G}{\hbar}\!\int_0^t\D s\!\int \D \rr\!\int\D\rr'\,\frac{f(t-s)}{|\rr-\rr'|}\\
     \times\comm{\hat \mu(\rr')}{\comm{e^{\tfrac i\hbar \hat H_N(s-t)}\hat \mu(\rr)e^{-\tfrac i\hbar \hat H_N(s-t)}}{\hat \rho_\text{\tiny CM}(t)}},
\end{aligned}
\end{equation}
and substitutes that in Eq.~\eqref{me}.
One recovers the standard DP model when $f(t-s)=\delta(t-s)$.
Along the lines of the previous analysis, we can neglect the Hamiltonian evolution and we find that  the CM density matrix evolves as in Eq.~\eqref{offdiagonal} with a time-dependent timescale $\tau(\db,t)$  reading
\begin{equation} \label{NMtimescale}
    \tau(\db,t)=\frac{\hbar}{\Delta E(\db)}\frac{t}{g(t)},
\end{equation}
where {$g(t)=2\int_0^t\D s\int_0^s\D s'\,f(s-s')$} and $\Delta E(\db)$ as in  Eq.~\eqref{DeltaEij}.

An already considered choice for the two-time correlation \cite{bassi2009non,carlesso2018colored} is $f(t)={\Omega_\text{C}}e^{-\Omega_\text{C}|t|}/2$, which leads to 
\begin{equation}\label{eq.2timecorr}
    {g(t>0)=t \left[1-\frac{1}{\Omega_\text{C}t}\left( 1-  e^{-\Omega_\text{C}t} \right)\right]},
\end{equation}
where the frequency cutoff $\Omega_\text{C}$ becomes a further parameter of the DP model. In the limit of $\Omega_\text{C}\to\infty$, one recovers the standard, white noise DP model, with {$g(t)=t$}. It is clear from Eq.~\eqref{eq.2timecorr} that one has $g(t)<t$ for all finite values of $\Omega_\text{C}>0$.
Consequently, the time of collapse in the colored version of the model, given by  $\tau(\db,t)$ in Eq.~\eqref{NMtimescale} is always larger than that of the standard DP model. Therefore, classicality takes longer to be achieved. Although one could make different choices of $f(t)$, the physically motivated ones should decay in time and we do not expect qualitative differences in the result.
\\

\textit{Conclusions. --} 
The DP model finds a unique place in our efforts to understand of the quantum/gravity interplay. It proposes a new path towards the formulation of a quantum theory of gravity, by assuming that the latter is responsible for the collapse of the  wave function. 
The model has been and is still subject to experimental verification, with different platforms. In this work we studied how effective the DP collapse is in predicting the emergence of a macroscopic classical world from an underlying quantum structure.  The conclusion is that not all macroscopic objects collapse effectively,  {meaning that there are some objects that do not collapse within the perception time of the human eye although they can be directly seen by it.} This implies that, in principle, we should see quantum effects (whatever that means) without the need of devices, like magnifying glasses, if  the macroscopic system is sufficiently well isolated. If its size increases by a couple of orders of magnitude in length or thickness, the collapse becomes fast enough.

Our analysis, therefore, shows that the quantum-to-classical transition  {occurs roughly at the border ($10^{10}-10^{12}$ atoms) between what we consider meso and macro}. In this region, the collapse is roughly independent from $R_0$ for a large range of values ($1-10^6$\,\AA). This is relevant since  the existence of $R_0$ seems artificial, and in fact, in the past, ways to removed it were proposed \cite{diosi1987universal,penrose2000,penrose2014gravitization}, which however were not successful \cite{ghirardi1990continuous,donadi2021underground}. It will be interesting to see whether this boundary set by the gravitational collapse of the wavefunction is more than a coincidence. 

To fully test this boundary up to $R_0=10^6\,$\AA, experiments should be considerably developed. The most promising experiments are the so-called non-interferometric  ones \cite{carlesso2022present}, which are based on detection of diffusive effects due to the collapse. To date, the strongest bound $R_0\gtrsim 4\,$\AA\;is set by experiments searching for collapse-induced radiation emission from a Germanium crystal \cite{arnquist2022search}. Since the predicted radiation emission rate is proportional to $R_0^{-3}$, an improvement of the bound on $R_0$ of six orders of magnitude (required to close the gap between experiments and theory) requires 18 orders of magnitude improvement on the sensitivity. While the expected  radiation emission rate can be increased by considering photons at lower energies (order of few keV) \cite{piscicchia2024x} rather than those considered in \cite{arnquist2022search}, it is not obvious how to fill such a large gap. Another type of experiments which might be relevant are  based on the detection of the spontaneous heating, which set a lower bound $R_0\gtrsim 4.6\times 10^{-2}$ \AA$\;$ \cite{vinante2021gravity}. This bound is two orders of magnitude lower than that from radiation, but it has the merit to be more robust to possible non-Markovian generalizations of the DP model. However, also in this case, the predicted increase is proportional to $R_0^{-3}$, requiring an improvement in sensitivity of 24 orders of magnitude to reach the upper bound.

\textit{Acknowledgments. --} The authors acknowledge Lajos Di\'osi for feedbacks on a early version of the manuscript. They acknowledge support from the University of Trieste, INFN, the PNRR MUR projects PE0000023-NQSTI and CN00000013-HPC, the EU EIC
Pathfinder project QuCoM (10032223), the UKRI through Grant 366
No.~EP/X021505/1 and FVG MICROGRANT LR2/2011 (D55-microgrants24).

\bibliography{Article_DP_theoretical_bound}

\newpage

\vfill{}\quad
\newpage

   \onecolumngrid
   \appendix

\section*{Appendix}

\section{Dynamical evolution of the center of mass of a system of $N$ particles in the Di\'osi-Penrose model}

\subsection{Di\'osi-Penrose equation for the center of mass of the system}

We start from the master equation of the DP model for $N$ particles, which is Eq.~\eqref{me} in the main text. Under the assumption of a rigid body, we can rewrite every position operator $\hat{\x}_i$ ($i=1,\dots,N$) of each $i$-th particle as 
\begin{equation}\label{xtoxclassical}
    \hat \x_i=\hat \x+\x_i^{(0)},
\end{equation}
where $\hat \x$ is the position operator of the center of mass (CM), and $\x_i^{(0)}$ is the classical displacement of the particle with respect to it.  We substitute Eq.~\eqref{xtoxclassical} in Eq.~\eqref{def.hat.mu} and merge the latter with Eq.~\eqref{me}, where for the sake of simplicity we take all $R_\text{eff,i}=R_\text{eff}$ to be equal. This gives
\begin{equation}
    \mathcal D\left[\hat \rho(t)\right]=-\frac{4\pi G}{\hbar}\int\D\rr \int\D\rr'\frac{1}{|\rr-\rr'|} \left[\hat \mu_{\text{\tiny CM}} (\rr'), [\hat \mu_{\text{\tiny CM}} (\rr),\hat \rho_{\text{\tiny CM}}] \right],
\end{equation}
where $\hat \mu_{\text{\tiny CM}} (\rr)$ is the mass density of the center of mass, which reads
\begin{equation}
    \hat \mu_{\text{\tiny CM}}(\rr)=\sum_{i=1}^N \frac{m_i}{(2\pi R_{\text{eff}}^2)^{3/2}}\exp\left[-\frac{(\hat \x+\x_i^{(0)}- \rr )^2}{2R_{\text{eff}}^2}\right].
\end{equation}
Then, by tracing over the internal degrees of freedom, and expressing the CM mass density in Fourier space
{\begin{equation}
    \hat \mu_\text{\tiny CM}(\rr)= \sum_{i=1}^N \frac{m_i}{(2\pi)^3} \int \D \q\, e^{-i\q\cdot(\hat \x+\x_i^{(0)}- \rr )} e^{-R_{\text{eff}}^2 q^2/2},
\end{equation}}
we find the master equation for the center of mass: $\tfrac{\D}{\D t}\hat \rho_\text{\tiny CM}(t)=-\tfrac i \hbar\comm{\hat H_\text{\tiny CM}}{\hat \rho_\text{\tiny CM}(t)}+\mathcal D[\hat \rho_\text{\tiny CM}(t)]$, where
\begin{equation}
\begin{aligned}
    &\mathcal D\left[\hat \rho_\text{\tiny CM}(t)\right]=\\
    &-\frac{4\pi G}{\hbar}\int\D\rr \int\D\rr'\frac{1}{|\rr-\rr'|}\sum_{i,j=1}^N\frac{m_im_j}{(2\pi)^6}\int\D \q\int\D \kk\, e^{-R_\text{eff}^2(\q^2+\kk^2)/2}e^{-i\q\cdot(\x_i^{(0)}-\rr')}e^{-i\kk\cdot(\x_j^{(0)}-\rr)}\comm{e^{-i\q\cdot\hat\x}}{\comm{e^{-i\kk\cdot\hat\x}}{\hat \rho(t)}}.
\end{aligned}
\end{equation}
We can simplify such an expression through the use of the following identity
\begin{equation}\label{eq.fourierNewton}
    \frac{1}{|\rr-\rr'|}=\frac{1}{2\pi^2}\int\D \p\, \frac{e^{-i\p\cdot(\rr-\rr')}}{p^2},
\end{equation}
and straightforward calculations lead to the following dynamical evolution of the center of mass of the system
\begin{equation}\label{MEDPA1}
    \frac{\D \hat \rho_\text{\tiny CM}(t)}{\D t}=-\frac{i}{\hbar}\comm{\hat H_\text{\tiny CM}}{\hat \rho_\text{\tiny CM}(t)}+\int\D\p\, F(\p)\left(e^{i\p\cdot\hat \x}\hat \rho_\text{\tiny CM}(t)e^{-i\p\cdot\hat \x}-\hat \rho_\text{\tiny CM}(t)\right),
\end{equation}
where we introduced 
\begin{equation}
    F(\p)= \frac{4G}{\pi\hbar}\sum_{i,j=1}^Nm_im_j\frac{e^{-R_\text{eff}^2p^2}}{p^2}e^{i\p\cdot(\x_i^{(0)}-\x_j^{(0)})},
\end{equation}
which is the form factor accounting for the dimension and form of the system. 
Equation \eqref{MEDPA1} is the master equation of the DP model for the center of mass of a $N$-particle system.

\subsection{Analytic solution of the equation for the center of mass}
To solve the master equation for the CM, it is convenient to introduce the characteristic function $\chi$ \cite{savage1985,smirne2010}, which is defined as
\begin{equation}\label{eq.def.chi}
    \chi(\nnu, \mmu, t)=\Tr{
    \hat \rho_\text{\tiny CM}(t)e^{\tfrac i\hbar(\nnu\cdot\hat\x+\mmu\cdot\hat \p)}}.
\end{equation}
For the sake of simplicity, we assume that the CM is free, i.e.~$\hat{H}_\text{\tiny CM}=\hat{\p}^2/2M$. Thus, 
given the master equation \eqref{MEDPA1} for a free particle, one finds that the characteristic function satisfies the following equation \cite{smirne2010}
\begin{equation}\label{chidiff}
    \frac{\D}{\D t}    \chi(\nnu, \mmu, t)=\frac{1}{M}\nnu\cdot\nabla_\mmu     \chi(\nnu, \mmu, t)+\frac{1}{\hbar}\left(E(\mmu)-E(\mathbf{0})\right)    \chi(\nnu, \mmu, t),
\end{equation}
where 
\begin{equation} \label{fDPdef}
    E(\mmu)=\hbar\int\D \p\, F(\p)e^{i\mmu\cdot \p},
\end{equation}
The solution to Eq.~\eqref{chidiff} is given by \cite{smirne2010}
\begin{equation}
    \chi(\nnu, \mmu, t)=\chi_0(\nnu, \mmu+\nnu \tfrac{t}{M}, t)\exp \left[\tfrac1\hbar{\int_0^t\D \tau \left(E(\mmu+\nnu \tfrac{\tau}{M})-E(\mathbf{0})\right)}\right],
\end{equation}
where $\chi_0(\nnu, \mmu, t)$ satisfies
\begin{equation}
    \frac{\D}{\D t}\chi_0(\nnu, \mmu, t)=\frac{1}{M}\nnu\cdot\nabla_\mmu \chi_0(\nnu, \mmu, t),
\end{equation}
and corresponds to the characteristic function for the Hamiltonian dynamics. This corresponds to the standard quantum mechanical predictions. 

To reconstruct the matrix elements of $\hat \rho_\text{\tiny CM}(t)$, one employs \cite{smirne2010}
\begin{equation} \label{rhoxy}
    \braket{\x|\hat \rho_\text{\tiny CM}(t)|\y}=\frac{1}{(2\pi\hbar)^3}\int\D\nnu\, e^{-\tfrac{i}{2\hbar}\nnu\cdot(\x+\y)}\chi(\nnu,\x-\y,t).
\end{equation}
We can express $\braket{\x|\hat \rho_\text{\tiny CM}(t)|\y}$ in terms of the matrix elements of $\hat{\rho}_\text{\tiny CM}^\text{\tiny QM}(t)$, which is the statistical operator corresponding to the quantum dynamical evolution (that related to $\chi_0$ alone). 
Such a relation reads
\begin{equation} \label{rhoDPxy}
\braket{\x|\hat{\rho}_\text{\tiny CM}(t)|\y}=\frac{1}{(2 \pi \hbar)^3} \int \D \nnu \int \D \z  \,e^{\frac{i}{\hbar}\nnu \cdot \z}\braket{\z+\x|\hat{\rho}_\text{\tiny CM}^\text{\tiny QM}(t)|\z+\y} \exp \left[\tfrac1\hbar\int_0^t \D \tau \left(E \left(\nnu \frac{\tau}{M}+\x-\y \right)-E(\mathbf{0})\right) \right],
\end{equation}
with $E(\x)$ being defined as in Eq.~\eqref{fDPdef} and which takes the explicit form of
\begin{equation} \label{fDP}
    E(\mmu)={8\pi G}\sum_{i,j=1}^Nm_im_j   \frac{{\erf}\left(\frac{|\x_i^{(0)}-\x_j^{(0)}+\mmu|}{2R_\text{eff}}\right)}{|\x_i^{(0)}-\x_j^{(0)}+\mmu|},
\end{equation}
for the DP model. Finally, the expression in Eq.~\eqref{DeltaEij}, namely
\begin{equation}\label{tutta}
\Delta E(\db) = 8 \pi G m^2 \sum_{i,j=1}^{N}
\left[ \frac{\erf \left( \frac{r_{ij}}{2 R_\text{eff}} \right)}{r_{ij}} -\frac{\erf \left( \frac{|\db+\rr_{ij}|}{2 R_\text{eff}} \right)}{|\db+\rr_{ij}|} \right],
\end{equation}
is obtained with $\Delta E(\db)=E(\db)-E(\mathbf 0)$.

\subsection{Neglecting the free evolution $\hat{H}_{\text{\tiny CM}}$ in the Di\'osi-Penrose master equation}

In the following, we investigate the conditions   under which one can safely neglect the free evolution (i.e.~$\hat H_\text{\tiny CM}$) when quantifying the collapse action of the DP model on the center of mass (CM).
Specifically, we focus on the matrix element of Eq.~\eqref{rhoDPxy} for the case of an initial state $\psi(\x,0)$ being 
in a superposition of two equal Gaussians centered in $\pm \db/2$. Hence, we have
\begin{equation}\label{eq.initial.superp}
\psi(\x,0)=\frac{1}{\mathcal{N}} \left[e^{-\frac{1}{2 \sigma^2}(\x-\db/2)^2} +e^{-\frac{1}{2 \sigma^2}(\x+\db/2)^2} \right],
\end{equation}
where
\begin{equation}
    \mathcal{N}=\left[2 (\sqrt{\pi}\sigma)^3 \left(1+e^{-\frac{d^2}{4\sigma^2}}\right) \right]^{1/2},
\end{equation}
is the corresponding normalization constant,
$d=|\db|$, {and $\sigma$ is the Gaussian spread}. The free, quantum mechanical  evolution of the CM, where $\hat H_\text{\tiny CM}=\hat{\p}^2/2M$, leads to the following expression for the state at a time $t$
\begin{equation} \label{psiQM}
\psi(\x,t)=\frac{1}{\mathcal{N}} \left(\frac{\sigma}{\sqrt{\sigma^2 + \frac{i t \hbar}{M}}} \right)^3 \left(e^{-\frac{(\x + \db/2)^2}{2 \left(\sigma^2 + \frac{i \hbar t}{M} \right)}}+e^{-\frac{(\x-\db/2)^2}{2 \left(\sigma^2 + \frac{i \hbar t}{M} \right)}} \right).
\end{equation}
We now focus on the matrix element $\braket{-\db/2|\hat{\rho}_\text{\tiny CM}(t)|\db/2}$, which measures the coherence between the two wave packets. Due to the spatial decoherence induced by the DP model, we expect this term to decay. Our goal is to understand under which conditions, when computing this term, $\hat{H}_\text{\tiny CM}$ can be neglected. 

From Eq.~\eqref{psiQM}, we can directly calculate the matrix element for the density operator in the standard quantum mechanical case as
{\begin{equation}  \label{eq.rho.cm.qm.t}\braket{\x|\hat{\rho}_\text{\tiny CM}^\text{\tiny QM}(t)|\y}=\psi^*(\x,t)\psi(\y,t)= \frac{1}{\mathcal{N}^2} \frac{\sigma ^6 \left(e^{-\frac{(\x + \db/2)^2}{2 \left(\sigma^2 - \frac{i \hbar t}{M} \right)}}+e^{-\frac{(\x-\db/2)^2}{2 \left(\sigma^2 - \frac{i \hbar t}{M} \right)}} \right) \left(e^{-\frac{(\y + \db/2)^2}{2 \left(\sigma^2 + \frac{i \hbar t}{M} \right)}}+e^{-\frac{(\y-\db/2)^2}{2 \left(\sigma^2 + \frac{i \hbar t}{M} \right)}} \right) }{\left(\sigma ^4 + ( \frac{h t}{M})^2\right)^{3/2}},
\end{equation}}
which can be then inserted in Eq.~\eqref{rhoDPxy}. After integrating over $\z$, we obtain
\begin{equation}\label{Kj40}
\braket{-\db/2|\hat{\rho}_\text{\tiny CM}(t)|\db/2}=\mathcal K_1 + \mathcal K_2 +\mathcal  K_3, \qquad \text{where} \qquad  
\mathcal K_j=\int \D \nnu\, e^{-\nnu^2}\mathcal{F}(\db/2,\nnu)\mathcal{G}_j(\db/2,\nnu).
\end{equation}
where we defined
\begin{equation} \label{contributions}
\begin{aligned}
\mathcal{G}_1(\db/2,\nnu)&=1,\\
\mathcal{G}_2(\db/2,\nnu)&=\exp\left(-\frac{d^2}{\sigma^2}\right)\exp\left(-\frac{2 \hbar t}{M \sigma^2}\frac{\nnu \cdot \db}{\sqrt{1+ \left(\frac{\hbar t}{M \sigma^2} \right)^2}} \right),\\
\mathcal{G}_3(\db/2,\nnu)&=\exp \left(-\frac{d^2}{4\sigma^2} \right)\exp\left(-\frac{\hbar t}{M \sigma^2}\frac{\nnu \cdot \db}{\sqrt{1+\left( \frac{\hbar t}{M \sigma^2}\right)^2}} \right)\times2 \cos \left(\frac{1}{\sigma}\frac{\nnu \cdot \db}{\sqrt{1+\left(\frac{\hbar t}{M \sigma^2} \right)^2}}  \right),\\
\mathcal{F}(\db/2,\nnu)&=\frac{1}{{\pi^{3/2} \mathcal{N}^2}\left[1+ \left(\frac{\hbar t}{M \sigma^2} \right)^2 \right]^{3/2}} \exp \left[\int_0^t \D \tau \left(f \left( \db + 2 \sigma \frac{\tau}{t}\frac{\nnu}{\sqrt{1+\left(\frac{M \sigma^2}{\hbar t} \right)^2}} \right)-f(\mathbf{0}) \right) \right]. 
\end{aligned}
\end{equation}
We notice that, the exponential factor in Eq.~\eqref{Kj40} limits the main contributions to $\mathcal K_j$ to the vectors $\nnu$ such that $|\nnu| \leq 1$.
Moreover, for $d \gg \sigma$, we can approximate the argument in the exponential of $\mathcal{F}(\db/2,\nnu)$ in Eq.~\eqref{contributions} as  $ (f(\db)-f(\mathbf{0}))t$, which is possible since $\nnu$ is limited. These two approximations allow to easily compute $\mathcal K_j$, which read
\begin{equation}\label{EQA18}
\begin{aligned}
\mathcal K_1&={\frac{1}{\mathcal{N}^2}}\left( \frac{1}{1+\left(\frac{\hbar t}{M \sigma^2} \right)^2} \right)^{3/2} \exp \left[ \left(f(\db)-f(\mathbf{0}) \right)t \right],\\
\mathcal K_2&= \exp \left[-\frac{d^2}{\sigma^2}\left(1-\frac{t^2 \hbar^2}{M^2 \sigma^4+ t^2 \hbar^2} \right) \right]\mathcal K_1,\\
\mathcal K_3&=2 \exp \left(-\frac{d^2}{4\sigma^2}\frac{2 M^2 \sigma^4}{M^2 \sigma^4 + t^2 \hbar^2} \right)\cos \left(\frac{M \sigma^2 t \hbar}{M^2 \sigma^4 + t^2 \hbar^2} \frac{d^2}{2 \sigma^2}\right)\mathcal K_1.
\end{aligned}
\end{equation}
For small times, satisfying $t \ll M \sigma^2/\hbar$, we see that the expressions for $\mathcal K_i$ reduce to
\begin{equation}
\begin{aligned}
\mathcal K_1 &= \frac{1}{\mathcal{N}^2} \exp[f(\db)-f(\mathbf{0}))t],\\
\mathcal K_2 &= e^{-\tfrac{d^2}{ \sigma^2}}\mathcal K_1 ,\\
\mathcal K_3 &= 2 e^{-\tfrac{d^2}{2 \sigma^2}} \mathcal K_1,
\end{aligned}
\end{equation}
and therefore 
\begin{equation} \label{CMH0}
\braket{-\db/2|\hat{\rho}_{\text{\tiny CM}}(t)|\db/2}=\frac{1}{\mathcal{N}^2} \left(1+ e^{-\tfrac{d^2}{2 \sigma^2}} \right)^2 \exp \left[(f(\db)-f(\mathbf{0}) )t\right],
\end{equation}
which is the expression that one would obtain if the evolution due to the Hamiltonian $\hat{H}_{\text{CM}}$ is neglected.
Indeed, when neglecting $\hat H_\text{CM}$, one has that Eq.~\eqref{chidiff} changes in 
\begin{equation}
    \frac{\D}{\D t}    \chi(\nnu, \mmu, t)=\left(f(\mmu)-f(\mathbf{0})\right)    \chi(\nnu, \mmu, t),
\end{equation}
whose trivial solution is 
\begin{equation}
    \chi(\nnu, \mmu, t)=\chi(\nnu, \mmu, 0)\exp[(f(\db)-f(\mathbf{0}))t],
\end{equation}
where $\chi(\nnu, \mmu, 0)$ can be obtained from Eq.~\eqref{eq.def.chi} by setting $t=0$.
Correspondingly, Eq.~\eqref{rhoDPxy} changes in 
\begin{equation} \label{rhoDPxyH0}
\braket{\x|\hat{\rho}_\text{\tiny CM}(t)|\y}=\braket{\x|\hat{\rho}_\text{\tiny CM}^\text{\tiny QM}(0)|\y} \exp[(f(\db)-f(\mathbf{0}))t],
\end{equation}
 where 
\begin{equation}
    \braket{\x|\hat{\rho}_\text{\tiny CM}^\text{\tiny QM}(0)|\y}=\psi^*(\x,0)\psi(\y,0).
\end{equation}
The trivial substitution of the latter expression in Eq.~\eqref{rhoDPxyH0}, and setting $\x=-\db/2$ and $\y=\db/2$ leads to Eq.~\eqref{CMH0}. Thus, under these assumptions, we can safely neglect the evolution due to $\hat{H}_\text{\tiny CM}$ in the Di\'osi-Penrose model.

\section{{Analytical} study of the behaviour of $ \Delta E ({\db})$}
\label{app_dE_behaviour}

We study the behaviour of $\Delta E(\db)$ for a crystal of $N$ atoms of mass $m$. We study a two-dimensional crystal with a monoatomic square lattice with the same lattice step $a$, and with crystal's sides of approximately the same length  $L$. Accordingly $L\sim \sqrt{N}a$. One can trivially generalize to the case of different length sides.
Specifically, we  analyze the case of a spatial superposition of the form in Eq.~\eqref{eq.initial.superp}, 
where the two branches of the superposition are well separated with respect to $L$. Hence, the length  hierarchy is given by $a\ll L\ll d$.
With this setting, the vectors $\rr_{ij}$ can only take discrete values, namely
$\rr_{ij}=a\left(n_x^i-n_x^j,n_y^i-n_y^j\right)$, where $n_k^i\in \mathbb Z$.

\begin{figure}[ht!] 
\centering
\includegraphics[width=0.65\textwidth]{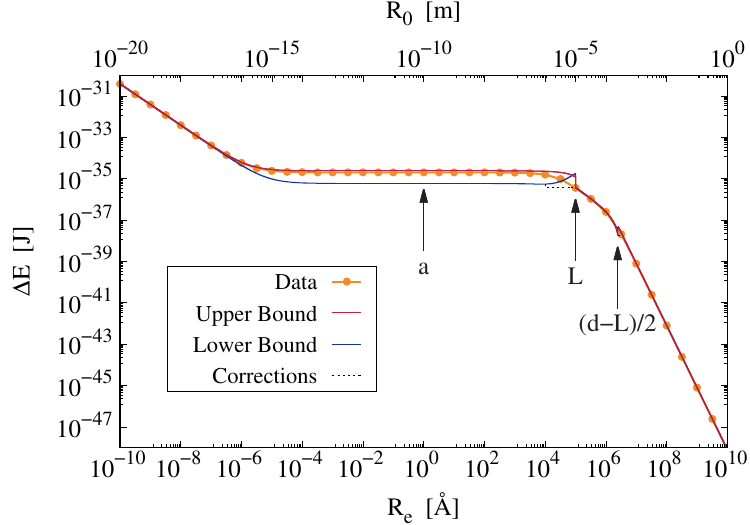} 
\caption{Analytical upper (red line) and lower (blue line) bounds for $\Delta E$ as a function of $R_\text{eff}$. The dashed black line provides a better estimation of the lower bound for values of $R_\text{eff}\simeq L$ and $(d-L)/2$ by exploiting the monotonicity of $\Delta E(\db)$. For comparison, we reported the exact values computed numerically (orange dots). The vertical arrows correspond  to the values of $R_\text{eff}$ equal  to $a$, $L$ and $(d-L)/2$ (from left to right).   {The values of the  parameters are: $a=1.0$\,\AA, $N = 10^{10}$, $L = 10^5$\,\AA \, and $d = 50 L = 5 \times 10^6$\,\AA.}}
\label{img_appendix_B_0}
\end{figure}

In the following, we analyze analytically the behaviour of $\Delta E(\db)$ in Eq.~\eqref{tutta} for different values of $R_\text{eff}$. Specifically, we provide the contributions to the sum from the first and second term in the square bracket. Namely, these are
\begin{equation}\label{def.S1S2}
    \begin{aligned}
        S_1&=\sum_{i,j=1}^{N}
 \frac{\erf \left( \frac{r_{ij}}{2 R_\text{eff}} \right)}{r_{ij}},\\
        S_2&=\sum_{i,j=1}^{N}\frac{\erf \left( \frac{|\db+\rr_{ij}|}{2 R_\text{eff}} \right)}{|\db+\rr_{ij}|},
    \end{aligned}
\end{equation}
so that $\Delta E(\db)=8\pi Gm^2 (S_1-S_2)$. The analytical expressions will be then compared with those obtained numerically in Fig.~\ref{img_appendix_B_0}.

\subsection{Analytical estimation of the first contribution to the sum $S_1$}

The first contribution to the sum $S_1$ is analysed in four different intervals of interest: $R_\text{eff}\leq a$, $a\leq R_\text{eff} \leq L$, $L\leq R_\text{eff}\leq d$ and $d\leq R_\text{eff}$.
We split the sum in $S_1$ in two contributions: the sum of the terms with $i=j$ and the sum where $i\neq j$. 

When $i=j$, we have $r_{ii}=0$ and thus the corresponding sum reads
\begin{equation}\label{eq.riito0}
    \sum_{i=1}^N\frac{\erf\left(\frac{r_{ii}}{2R_\text{eff}}\right)}{r_{ii}}\underset{r_{ii}\to0}{\longrightarrow}\sum_{i=1}^N\frac{1}{\sqrt{\pi}R_\text{eff}},
\end{equation}
and we have
\begin{equation}\label{eq.S1}
    S_1=\frac{N}{\sqrt{\pi}R_\text{eff}}+\sum_{\substack{i,j=1\\{i\neq j}}}^{N}
 \frac{\erf \left( \frac{r_{ij}}{2 R_\text{eff}} \right)}{r_{ij}},
\end{equation}
which we will study below.

\subsubsection{First interval: $R_\text{eff} \leq a$} 
Consider the case $R_\text{eff} \leq a$. 
For all terms with $i\neq j$, 
we have that $r_{ij}\geq a$. Indeed, the minimal value of $r_{ij}=a$ {is attained for the} first nearest neighbor atoms.
This implies $R_\text{eff}\leq r_{ij}$. In particular, we can divide the contributions to the sum in two cases
\begin{equation}
    \begin{cases}
    \text{for}\ 1\leq\frac{r_{ij}}{R_\text{eff}}\leq 4,&\tfrac12<\erf \left( \frac{r_{ij}}{2 R_\text{eff}} \right)
        <1,\\
    \text{for}\ 4\leq\frac{r_{ij}}{R_\text{eff}},&\erf \left( \frac{r_{ij}}{2 R_\text{eff}} \right)
        \sim1.
    \end{cases}
\end{equation}
We can then employ these to set an upper and lower bound on the sum by approximating that all the terms weight the same independently of the value of $r_{ij}$. We have then
\begin{equation}\label{biounds1}
    \sum_{\substack{i,j=1\\{i\neq j}}}^{N}
 \frac12\frac{1}{r_{ij}}\leq\sum_{\substack{i,j=1\\{i\neq j}}}^{N}
 \frac{\erf \left( \frac{r_{ij}}{2 R_\text{eff}} \right)}{r_{ij}}\leq\sum_{\substack{i,j=1\\{i\neq j}}}^{N}
 \frac{1}{r_{ij}}.
\end{equation}
Then, both limits depend on the same sum, which can be rewritten as
\begin{equation}
    \sum_{\substack{i,j=1\\{i\neq j}}}^{N}
\frac{1}{r_{ij}}=\sum_{\substack{i,j=1\\{i\neq j}}}^{N}\frac{1}{a\sqrt{(n_x^i-n_x^j)^2+(n_y^i-n_y^j)^2}},
\end{equation}
where $n_k^i\in[-\sqrt{N}/2,\sqrt{N}/2]$. We first fix the $i$-th particle as that at the center of the lattice with $n_x^i=n_y^i=0$ and sum over $j$. This term provides an upper bound on the other contributions of the $i$-sum, since fixing the $i$-th atom to any other in the lattice would provide a smaller contribution. Then, the corresponding vector $\rr_{ij}=a(-n_x^j,-n_y^j)$ has a length of $r_{ij}=a\sqrt{(n_x^j)^2+(n_y^j)^2}$. We obtain
\begin{equation}\label{C3}
\sum_{\substack{i,j=1\\{i\neq j}}}^{N} \frac{1}{r_{ij}} \leq \sum_{i=1}^{N}  \underset{(n^j_{x},n^j_{y})\neq(0,0)}{\sum_{n^j_{x},n^j_{y}=-\sqrt{N}/2}^{\sqrt{N}/2}}\frac{1}{a \sqrt{(n^j_{x})^{2}+(n^j_{y})^{2}}} = \frac{N}{a}  \underset{(n^j_{x},n^j_{y})\neq(0,0)}{\sum_{n^j_{x},n^j_{y}=-\sqrt{N}/2}^{\sqrt{N}/2}}\frac{1}{\sqrt{(n^j_{x})^{2}+(n^j_{y})^{2}}}.
\end{equation}
The contribution in right hand side of Eq.~\eqref{C3} can be approximated with the following integral
\begin{equation}
   \frac{N}{a} \int_{-\sqrt{N}/2}^{\sqrt{N}/2}\D x    \int_{-\sqrt{N}/2}^{\sqrt{N}/2}\D y\,\frac{1}{\sqrt{x^2+y^2}}=\frac{\eta_+ N\sqrt{N}}{a},
\end{equation}
where $\eta_+=2 \ln(3+2\sqrt{2})\simeq3.5$.

 Similarly, if we fix the $i$-th atom to the corner of the lattice, we provide a lower bound. The corresponding vector $\rr_{ij}=a(-\sqrt{N}/2-n_x^j,-\sqrt{N}/2-n_y^j)$ has a length of $r_{ij}=a\sqrt{(\sqrt{N}/2+n_x^j)^2+(\sqrt{N}/2+n_y^j)^2}$. Thus, we have
\begin{equation}\label{C4}
\sum_{\substack{i,j=1\\{i\neq j}}}^{N} \frac{1}{r_{ij}}\geq     \sum_{i=1}^{N}  \underset{(n^j_{x},n^j_{y})\neq(0,0)}{\sum_{n^j_{x},n^j_{y}=-\sqrt{N}/2}^{\sqrt{N}/2}}\frac{1}{a \sqrt{(\sqrt{N}/2+n_x^j)^2+(\sqrt{N}/2+n_y^j)^2}}.
\end{equation}
Similarly as done before, we can approximate the sum to an integral, which reads
\begin{equation}
   \frac{N}{a} \int_{-\sqrt{N}/2}^{\sqrt{N}/2}\D x    \int_{-\sqrt{N}/2}^{\sqrt{N}/2}\D y\,\frac{1}{\sqrt{(\sqrt{N}/2+x)^2+(\sqrt{N}/2+y)^2}}=\frac{2\eta_- N\sqrt{N}}{a},
\end{equation}
where $\eta_-= \operatorname{arcsinh}(1)\simeq0.9$. By merging the inequality in Eq.~\eqref{biounds1} with those in Eq.~\eqref{C3} and Eq.~\eqref{C4}, we find the bounds on Eq.~\eqref{eq.S1} being
\begin{equation}\label{firstcontribution-firstinterval}
   \frac{N}{\sqrt{\pi}R_\text{eff}}+ \frac{\eta_- N\sqrt{N}}{a}\leq S_1\leq \frac{N}{\sqrt{\pi}R_\text{eff}}+\frac{\eta_+ N\sqrt{N}}{a}.
\end{equation}
Specifically, for
\begin{equation}
    R_\text{eff}\lesssim \frac{a}{\eta_\pm \sqrt{\pi N}},
\end{equation}
the first contribution is stronger and corresponds to a behaviour going as $\sim R_\text{eff}^{-1}$. Conversely, for larger values of $R_\text{eff}$ (but still within the first interval, $R_\text{eff}<a$), the first term in Eq.~\eqref{firstcontribution-firstinterval} is negligible with respect to the second one, which is independent from $R_\text{eff}$.

\subsubsection{Second interval: $a \leq R_\text{eff} \leq L$} 

In the second interval, $a \leq R_\text{eff} \leq L$, we have that $r_{ij}$ can be comparable with $R_\text{eff}$.  Namely, for $i\neq j$, we have $a\leq r_{ij}\leq\sqrt{2}L$. Due to the comparable interval of $r_{ij}$ and $R_\text{eff}$, we have several edge effects that make the analytical estimation more difficult. Since one can approximate fairly enough the error function to 
\begin{equation}\label{eq.erf.cases}
    \erf\left(\frac{x}{2}\right)\simeq
    \begin{cases}
       \frac{x}{\sqrt{\pi}},&x\leq2,\\
       1,&x\geq2,
    \end{cases}
\end{equation}
we divide the sum in terms for which $r_{ij}\leq 2R_\text{eff}$ and terms for which  $r_{ij}\geq 2R_\text{eff}$.

For $r_{ij}\leq 2 R_\text{eff}$,
\begin{equation}
    \frac{\erf \left( \frac{r_{ij}}{2 R_\text{eff}} \right)}{r_{ij}}\simeq \frac{1}{\sqrt{\pi}R_\text{eff}},
\end{equation}
and, given the $i$-th atom, there are roughly $(2R_\text{eff}/a)^2$  atoms $j$ that are at most at a distance $2R_\text{eff}$. Thus, the contribution to the second term in Eq.~\eqref{eq.S1} for $r_{ij}\leq 2R_\text{eff}$ can be evaluated to
\begin{equation}\label{eq.rij<2Re}
    \sum_{\substack{i,j=1\\{i\neq j,\,r_{ij}\leq 2R_\text{eff}}}}^{N} \frac{\erf \left( \frac{r_{ij}}{2 R_\text{eff}} \right)}{r_{ij}}\simeq N\times \left(\frac{2R_\text{eff}}{a}\right)^2 \frac{1}{\sqrt{\pi}R_\text{eff}}=4N \frac{R_\text{eff}}{\sqrt{\pi}a^{2}}.
\end{equation}
For $r_{ij}\geq 2 R_\text{eff}$,
\begin{equation}
    \frac{\erf \left( \frac{r_{ij}}{2 R_\text{eff}} \right)}{r_{ij}}\simeq \frac{1}{r_{ij}}.
\end{equation}
Then, the corresponding contribution to the sum can be rewritten as
\begin{equation}\label{eq.rij>2Re}
    \sum_{\substack{i,j=1\\{i\neq j,\,r_{ij}\geq 2R_\text{eff}}}}^{N} \frac{\erf \left( \frac{r_{ij}}{2 R_\text{eff}} \right)}{r_{ij}}\simeq\sum_{\substack{i,j=1\\{i\neq j}}}^{N}\frac{1}{r_{ij}}-\sum_{\substack{i,j=1\\{i\neq j,\,r_{ij}\leq 2R_\text{eff}}}}^{N}\frac{1}{r_{ij}},
\end{equation}
where the upper and lower bounds to the first sum are  respectively given by Eq.~\eqref{C3} and Eq.~\eqref{C4}. 
Also the second term can be bounded. A lower bound is given by setting all the $r_{ij}$ to the maximum distance being $2 R_\text{eff}$, namely
\begin{equation}
    \sum_{\substack{i,j=1\\{i\neq j,\,r_{ij}\leq 2R_\text{eff}}}}^{N}\frac{1}{r_{ij}}\geq \sum_{\substack{i,j=1\\{i\neq j,\,r_{ij}\leq 2R_\text{eff}}}}^{N}\frac{1}{2R_\text{eff}}=N\times \left(\frac{2R_\text{eff}}{a}\right)^2\frac{1}{2R_\text{eff}}=\frac{4N R_\text{eff}}{a^2}.
\end{equation}
An upper bound is instead obtained by fixing the $i$-th atom to the center of the lattice, and approximating the sum over $j$ to an integral in spherical coordinates. Namely,
\begin{equation}
    \sum_{\substack{i,j=1\\{i\neq j,\,r_{ij}\leq 2R_\text{eff}}}}^{N}\frac{1}{r_{ij}}\leq \frac{2\pi N}{a} \mathcal F_G \int_1^{\tfrac{2R_\text{eff}}{a}}\D r =\frac{2\pi N}{a}\mathcal F_G \left(\frac{2R_\text{eff}}{a}-1\right),
\end{equation}
where $\mathcal F_G$ is a geometrical factor accounting for couples of atoms $(i,j)$ that are counted in the sum but do not appertain to the lattice (indeed, one can have that $2R_\text{eff}=2L\geq \sqrt{2}L$ being the maximum distance $r_{ij}$), and accounts for the ratio of the area of a square lattice of length  $2R_\text{eff}$ with the spam of the considered $r_{ij}$, namely that of a planar toric lattice of internal radius $a$ and external radius $2R_\text{eff}$:
\begin{equation}
    \mathcal F_G=\frac{L^2}{L^2+4LR_\text{eff}+\pi R_\text{eff}^2}\frac{4R_\text{eff}^2}{\pi(4R_\text{eff}^2-a^2)}.
\end{equation}
By summing the contributions from Eq.~\eqref{eq.rij<2Re} and Eq.~\eqref{eq.rij>2Re}, we find that Eq.~\eqref{eq.S1} is bounded as follows
\begin{equation}\label{firstcontribution-secondinterval}
   \frac{N}{\sqrt{\pi}R_\text{eff}} +\frac{\eta_-N\sqrt{N}}{a}+\frac{4NR_\text{eff}}{\sqrt{\pi}a^2}-\frac{2\pi N}{a}\mathcal F_G \left(\frac{2R_\text{eff}}{a}-1\right)\leq S_1\leq \frac{N}{\sqrt{\pi}R_\text{eff}} +\frac{\eta_+N\sqrt{N}}{a}+\frac{4NR_\text{eff}}{a^2}\left(\frac{1}{\sqrt{\pi}}-1\right).
\end{equation}
Specifically, for small values of $R_\text{eff}\gtrsim a$ the main contribution comes from the second term, i.e.~$\eta_\pm N\sqrt{N}/a$,  being independent from $R_\text{eff}$. For large values of $R_\text{eff}\lesssim L$, the third and fourth terms become the dominant ones. In particular, a change in the $R_\text{eff}$ dependence of $S_1$ appears from values of
\begin{equation}
    R_\text{eff}\gtrsim\frac{\eta_\pm N\sqrt{N\pi}a}{4}.
\end{equation}

\subsubsection{Third interval: $L \leq R_\text{eff} \leq d$} 

In this regime $r_{ij}\leq R_\text{eff}$ for $r_{ij}\leq L$. However, one also has distances such that $L\leq r_{ij}\leq \sqrt{2}L$ for which one can have $r_{ij}\geq R_\text{eff}$.
Nevertheless, we have that $\sqrt{2}L<2L\leq 2R_\text{eff}$, which implies that $r_{ij}\leq \sqrt{2}L<2R_\text{eff}$. Consequently, we can employ the small distance approximation in Eq.~\eqref{eq.erf.cases} for all $r_{ij}$, and Eq.~\eqref{eq.S1} becomes
\begin{equation}
    S_1\simeq\frac{N}{\sqrt{\pi}R_\text{eff}}+\sum_{\substack{i,j=1\\{i\neq j}}}^{N}
 \frac{1}{\sqrt{\pi}R_\text{eff}},
\end{equation}
where there are $N(N-1)$ terms in the sum. This gives
\begin{equation}\label{firstcontribution-thirdinterval}
    S_1\simeq\frac{N^2}{\sqrt{\pi}R_\text{eff}},
\end{equation}
whose behaviour with respect to $R_\text{eff}$ is the same in the entire interval.

\subsubsection{Fourth interval: $d\leq R_\text{eff}$} 
For the last interval, similar considerations are applied and the terms of the sum are approximated to their small distance limit. Consequently, we obtain
\begin{equation}\label{firstcontribution-fourthinterval}
    S_1\simeq\frac{N^2}{\sqrt{\pi}R_\text{eff}},
\end{equation}
exactly as in the third interval.

\subsection{Analytical estimation of the second contribution to the sum $S_2$}

For the second contribution to the sum [cf.~Eq.~\eqref{def.S1S2}], one needs to compare $R_\text{eff}$ with $|\db+\rr_{ij}|$. Considering that $\db$ is orthogonal  to one of the sides of the cubic lattice, we have the following bounds on $|\db+\rr_{ij}|$:
\begin{equation}\label{limitsdrij}
    d-L\leq |\db+\rr_{ij}|\leq \sqrt{(d+L)^2+L^2}\leq d+2L,
\end{equation}
The second contribution is  analysed in three different intervals of interest: $R_\text{eff}\leq (d-L)/2$, $(d-L)/2\leq R_\text{eff}\leq (d+2L)/2$ and $(d+2L)/2\leq R_\text{eff}$.

\subsubsection{First interval: $R_\text{eff}\leq (d-L)/2$}

In this interval, we have $R_\text{eff}\leq (d-L)/2\leq |\db+\rr_{ij}|/2$. Thus, 
we can safely apply the long distance approximation in Eq.~\eqref{eq.erf.cases} to $S_2$ in Eq.~\eqref{def.S1S2}, thus obtaining
\begin{equation}
    S_2\simeq\sum_{i,j=1}^{N}\frac{1}{|\db+\rr_{ij}|}.
\end{equation}
By applying the bounds in Eq.~\eqref{limitsdrij}, we can derive an upper and lower bound to $S_2$, which read
\begin{equation}\label{secondcontribution-firstinterval}
    \frac{N^2}{d+2L}\leq S_2\leq \frac{N^2}{d-L},
\end{equation}
where the $N^2$ factor follows trivially.

\subsubsection{Second interval: $(d-L)/2\leq R_\text{eff}\leq (d+2L)/2$}
In the second interval, $(d-L)/2\leq R_\text{eff}\leq (d+2L)/2$. Thus, by combining the definition of the interval with Eq.~\eqref{limitsdrij}, we obtain
\begin{equation}
    \frac{d-L}{d+2L}\leq\frac{|\db+\rr_{ij}|}{2R_\text{eff}}\leq\frac{d+2L}{d-L}.
\end{equation}
Since the terms of the sum in $S_2$ are monotonically decreasing with respect to $\tfrac{|\db+\rr_{ij}|}{2R_\text{eff}}$, for each of them we have
\begin{equation}
   \frac{1}{2R_\text{eff}} \frac{\erf \left( \frac{d+2L}{d-L} \right)}{\frac{d+2L}{d-L}}\leq\frac{\erf \left( \frac{|\db+\rr_{ij}|}{2 R_\text{eff}} \right)}{{|\db+\rr_{ij}|}}\leq\frac{1}{2R_\text{eff}}\frac{\erf \left( \frac{d-L}{d+2L}\right)}{\frac{d-L}{d+2L}}.
\end{equation}
It follows that
\begin{equation}\label{secondcontribution-secondinterval}
    \frac{\epsilon_-N^2}{2R_\text{eff}}\leq S_2\leq\frac{\epsilon_+N^2}{2R_\text{eff}},
\end{equation}
where
\begin{equation}
    \epsilon_\pm=\frac{\erf \left[\left( \frac{d- L}{d+2 L}\right)^{\pm 1}\right]}{\left(\frac{d- L}{d+2 L}\right)^{\pm1}},
\end{equation}
which can be quantified once fixed the ratio between $L$ and $d$. For example, for $d=4L$, we have $\epsilon_-\simeq0.5$ and $\epsilon_+\simeq 1$.

\subsubsection{Third interval: $(d+2L)/2\leq R_\text{eff}$}

In this interval, $R_\text{eff}\geq (d+2L)/2\geq |\db+\rr_{ij}|/2$. So we can apply the short distance approximation in Eq.~\eqref{eq.erf.cases}, which trivially gives
\begin{equation}\label{secondcontribution-thirdinterval}
    S_2\simeq \frac{N^2}{\sqrt{\pi}R_\text{eff}}.
\end{equation}

\subsection{Both contributions}

The sum of the two contributions to $\Delta E(\db)=8\pi Gm^2 (S_1-S_2)$ can be computed case by case depending on the value of $R_\text{eff}$ compared to the other  lengths involved. There are a total of six different intervals. Notably, when bounds are involved, one should account that $\min(\Delta E(\db))=8\pi Gm^2 (\min (S_1)-\max (S_2))$ and  $\max (\Delta E(\db))=8\pi Gm^2 (\max (S_1)-\min( S_2))$. The intervals and the corresponding equations to be used are reported in Tab.~\ref{tab.allDeltaE}.
In particular, for $(d+2L)/2\leq R_\text{eff}\leq d$ and $d\leq R_\text{eff}$, we have that the dominant contributions cancel, namely $S_1-S_2\simeq0$. For these intervals, one needs to consider also second order terms of the expansion. These give
\begin{equation}
    \Delta E (\db) \simeq 8 \pi G m^2 \sum_{i,j=1}^{N} \left(  - \frac{r^2_{ij}}{12 \sqrt{\pi} R_\text{eff}^3}+ \frac{|\db + \rr_{ij}|^2}{12 \sqrt{\pi} R_\text{eff}^3} \right)=\frac{2}{3} \sqrt{\pi} G m^2 \sum_{i,j=1}^{N} \frac{d^2 + 2 \  \rr_{ij} \cdot \db}{R_\text{eff}^3}.
\end{equation}
However, since $\rr_{ij}=-\rr_{ji}$ for all $i$ and $j$, then $\sum_{i,j=1}^N\rr_{ij}=0$, and one obtains
\begin{equation}\label{lastinterval}
    \Delta E(\db)\simeq\frac{2}{3} \sqrt{\pi} G m^2 N^2 \frac{d^2}{R_\text{eff}^3}.
\end{equation}

Table \ref{tab.allDeltaE} summarises the analytical study of $\Delta E(\db)$, which is graphically reported in Fig.~\ref{img_appendix_B_0} where it was compared with the numerical evaluation of the sum.
Finally, a last comment is due: $\Delta E(\db)$ is a monotonically decreasing function of $R_\text{eff}$, and this can be employed to estimate its values in those intervals where an analytical estimation is hard to be obtained. In particular, we exploited this for $R_\text{eff}$ just below $L$ and around $(d-L)/2$, to approximate the behaviour of $\Delta E(\db)$ with the black dashed line in Fig.~\ref{img_appendix_B_0}.

\renewcommand{\arraystretch}{1.5}
\begin{table}[t]
    \centering
    \begin{tabular}{c|c}
       Interval for $R_\text{eff}$&  Value or bounds on $\Delta E(\db)/(8\pi Gm^2)=S_1-S_2=\Delta S$\\
       \hline
 $R_\text{eff}\leq a$&$\frac{N}{\sqrt{\pi}R_\text{eff}}+ \frac{\eta_- N\sqrt{N}}{a}-\frac{N^2}{d-L}\leq \Delta S\leq \frac{N}{\sqrt{\pi}R_\text{eff}}+\frac{\eta_+ N\sqrt{N}}{a}-\frac{N^2}{d+L}$\\
$a\leq R_\text{eff}\leq L$&$\frac{N}{\sqrt{\pi}R_\text{eff}} +\frac{\eta_-N\sqrt{N}}{a}-\frac{N^2}{d-L}+\frac{4NR_\text{eff}}{\sqrt{\pi}a^2}-\frac{2\pi N}{a}\mathcal F_G \left(\frac{2R_\text{eff}}{a}-1\right)\leq \Delta S\leq \frac{N}{\sqrt{\pi}R_\text{eff}} +\frac{\eta_+N\sqrt{N}}{a}-\frac{N^2}{d+L}+\frac{4NR_\text{eff}}{a^2}\left(\frac{1}{\sqrt{\pi}}-1\right)$\\
$L\leq R_\text{eff}\leq \frac{(d-L)}2$&$\frac{N^2}{\sqrt{\pi}R_\text{eff}}-\frac{N^2}{d-L}\leq\Delta S\leq \frac{N^2}{\sqrt{\pi}R_\text{eff}}-\frac{N^2}{d+2L}$\\
$\frac{(d-L)}2\leq R_\text{eff}\leq \frac{(d+2L)}2$&$\frac{N^2}{\sqrt{\pi}R_\text{eff}}-\frac{\epsilon_+N^2}{2R_\text{eff}}\leq \Delta S\leq \frac{N^2}{\sqrt{\pi}R_\text{eff}}-\frac{\epsilon_-N^2}{2R_\text{eff}}$\\
$\frac{(d+2L)}2\leq R_\text{eff}\leq d$& $\Delta S=\frac{N^2d^2}{12\sqrt{\pi}R_\text{eff}^3}$\\
$d\leq R_\text{eff}$&   $\Delta S=\frac{N^2d^2}{12\sqrt{\pi}R_\text{eff}^3}$\\
    \end{tabular}
    \caption{Summary of the results of the analytical study of the $R_\text{eff}$ dependence of $\Delta E(\db)$ for different values of $R_\text{eff}$.}
    \label{tab.allDeltaE}
\end{table}

\section{Plateau dependence on the dimensionality}

\begin{figure}[t] 
\centering
\includegraphics[width=0.65\textwidth]{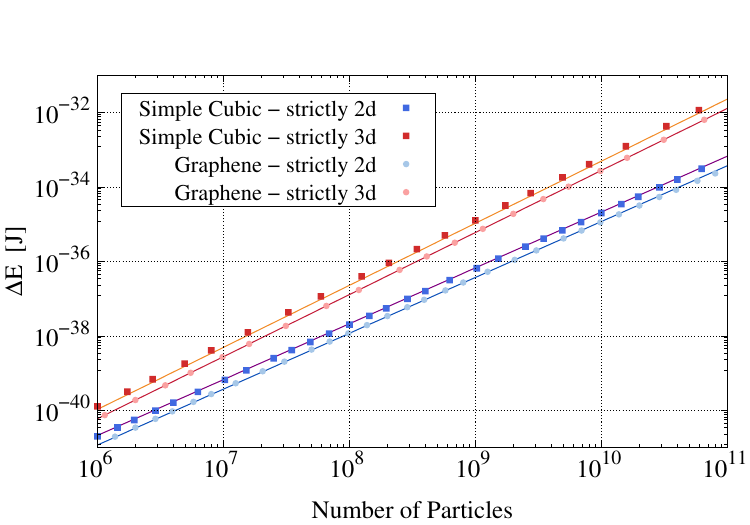} 
\caption{$\Delta E$ as a function of $N$ for different two- and three-dimensional systems and comparison with the analytical behaviours. We numerically computed and compared the two- and three-dimensional cases. For the two-dimensional case, we took a  sheet of graphene (blue dots) and a sheet of a material with a simple cubic lattice (blue squares). For the three dimensional case, we considered a cubic stack of graphene sheets (red dots) and a cube of material with a simple cubic lattice (red squares). The corresponding lines are obtained from Eq.~\eqref{eq_dE_vs_dimensionality} for the simple cubic lattice, while are instead fitted for the case of graphene with the same dependence from $N$ as in Eq.~\eqref{eq_dE_vs_dimensionality}. 
{The values of the  parameters are: $a=1.0$\,\AA, $d = 5 \times 10^6$\,\AA \ and $m = 2.0 \times 10^{-26}\,$kg for the simple cubic data and $d = 1 \times 10^6$\,\AA \ for the graphene one. In all cases, we fixed $R_\text{eff} = 1$\,\AA.}}
\label{img_appendix_B_2}
\end{figure}

Here, we quantify the dimensionality dependence of the plateau of $\tau(\db)$, and thus $\Delta E(\db)$,  for $R_\text{eff}\sim a$. 
For the sake of simplicity, we assume our system to be a sphere in $D\geq2$ dimensions. Namely, this corresponds to a disk for $D=2$ and a sphere for $D=3$ both of radius $L/2$.      
For $R_\text{eff}\sim a$,
$S_2$ can be safely bounded with Eq.~\eqref{secondcontribution-firstinterval}, which is independent from the dimensionality of the problem and from $R_\text{eff}$ as expected for the plateau.
Conversely, from Eq.~\eqref{eq.S1} we have
\begin{equation}
S_1=\frac{N}{\sqrt{\pi}a}+\sum_{\substack{i,j=1\\{i\neq j}}}^{N}\frac{\erf\left(\frac{r_{ij}}{2a}\right)}{r_{ij}}.
\end{equation}
Since, independently from the dimensionality of the problem, Eq.~\eqref{biounds1} still holds,  one only needs to evaluate 
\begin{equation}
    \sum_{\substack{i,j=1\\{i\neq j}}}^{N}\frac{1}{r_{ij}}=\sum_{\substack{i,j=1\\{i\neq j}}}^{N}\frac{1}{a\left(
   \sum_{k=1}^D(n_k^i-n_k^j)^2 
    \right)^{\tfrac12}}.
\end{equation}
By following the reasoning described above, we estimate the latter with its upper bound (see Eq.~\eqref{C3}, which provided a very good estimate for $S_1$ for $R_\text{eff}\simeq a$):
\begin{equation}
\sum_{\substack{i,j=1\\{i\neq j}}}^{N} \frac{1}{r_{ij}} \lesssim  \frac{N}{a}  \sum_{\{n_k^j\}_{k=1}^D\in S_D}\frac{1}{\left(
   \sum_{k=1}^D(n_k^j)^2 
    \right)^{\tfrac12}}.
\end{equation}
This is  the generalization of Eq.~\eqref{C3} to $D$ dimensions, with $S_D=\set{n_k^j\mid 0<|a(n_1^j,\dots,n_D^j)|<{L}/{2}}$ identifying a $D$-dimensional sphere on which the lattice is defined.
Such an expression can be simply approximated to a $1/r$ integral in $D$ dimensions, i.e.
\begin{equation}
   \frac{N}{a} \Omega_D\int_0^{N_D/2}\D x\,x^{D-1}\frac{1}{x}=\frac{N}{a}\frac{2 \pi^{D/2}}{\Gamma(D/2)}\frac{N^{\frac{D-1}{D}}}{2^{D-1}(D-1)},
\end{equation}
where $\Omega_D={2 \pi^{D/2}}/{\Gamma(D/2)}$ is the $D$-dimensional full angle. Correspondingly, for $R_\text{eff}\simeq a$, one obtains
\begin{equation}
    \Delta E(\db)\simeq  \frac{8\pi G m^2}{a}\frac{ \pi^{D/2}}{\Gamma(D/2)}\frac{N^{\frac{2D-1}{D}}}{2^{D-2}(D-1)},
\label{eq_dE_vs_dimensionality}
\end{equation}
which is plotted in Fig.~\ref{img_appendix_B_2} with the black and orange lines  for $D=3$ and $D=2$ dimensions respectively. For comparison, we report the numerical estimations of $\Delta E(\db)$.

\section{Algorithm for the numerical calculation}

The evaluation of Eq.~\eqref{DeltaEij} involves the calculation of the term
\begin{equation}\label{eq_basic_sum}
\sum_{i=1}^N \sum_{j=1}^N f(\rr_{ij}, R_0, \mathbf{d})  ,
\end{equation}
where $i$ and $j$ run across all the $N$ lattice sites and $f(\rr_{ij}, R_0, \mathbf{d})$ is a function that depends on the distance $\rr_{ij}$ between the sites $i$ and $j$.
In this form, the evaluation of Eq.~(\ref{eq_basic_sum}) requires the sum of $N^2$ terms.
Given the large systems considered in this work ($N \sim 10^{10}$), a direct evaluation of this sum is not feasible on a personal computer or even on a dedicated computer facility.
However, since the summed terms in Eq.~\eqref{eq_basic_sum} depend only on the distance between the sites, we can rewrite such an  equation as a single weighted sum across the lattice sites
\begin{equation}\label{eq_basic_fancy_sum}
\sum_{i=1}^N \sum_{j=1}^N f(\rr_{ij}, R_0, \mathbf{d}) = \sum_{\rr \in \mathcal{D}} \omega(\rr) f(\rr, R_0, \mathbf{d})  ,
\end{equation}
where $\mathcal{D}$ is the set of every possible distances $\rr$ between the lattice sites, and the weight $\omega(\rr)$ is the number of pairs of lattice cells that are at a distance $\rr$ apart.

\begin{figure}[h!] 
\centering
\includegraphics[width=0.5\textwidth]{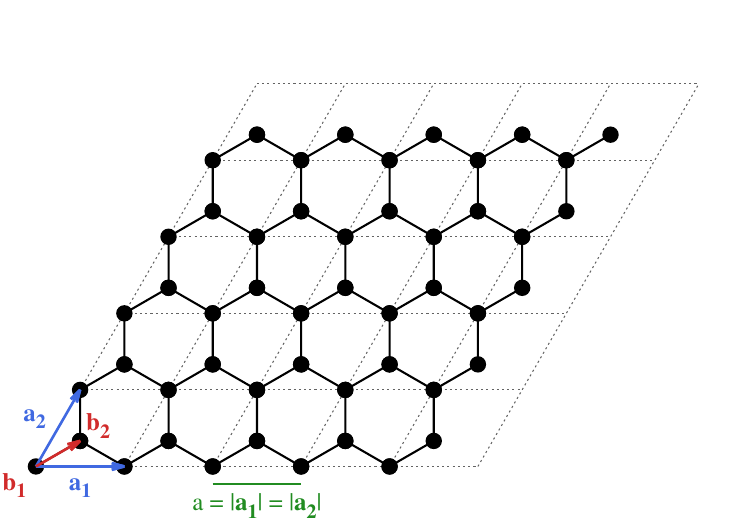} 
\caption{Graphical representation of the graphene lattice that was employed for the numerical summation. The lattice vectors are $\ab_1=a(1,0)$ and $\ab_2=a(\tfrac12,\tfrac{\sqrt{3}}{6})$. For each cell, we have two carbon atoms, which are identified by $\bb_1=(0,0)$ and $\bb_2=a(\tfrac12,\tfrac{\sqrt{3}}{6})$. }
\label{img_lattice}
\end{figure}

We consider a general two-dimensional lattice, with primitive vectors $\ba_1$ and $\ba_2$, {and basis $\bb_\alpha$, with $\alpha= 1,\dots,K$ being the index of the particles in the unit cell}.
{For instance, the graphene is a two-dimensional lattice with primitive vectors $\ba_1 = a \left( 1, 0 \right)$, $\ba_2 = a \left( \frac{1}{2}, \frac{\sqrt{3}}{2} \right)$, and basis $\bb_1 = \left( 0, 0 \right)$, $\bb_2 = a \left( \frac{1}{2}, \frac{\sqrt{3}}{6} \right)$, where $a=2.46$\,\AA\, is the lattice step~\cite{yang_structure_2018}.}
Each site of the lattice can be written as a linear combination of the primitive vectors with integer coefficients, ${\x_{i{,\alpha}}} = n_1^i \ba_1 + n_2^i \ba_2 {+\bb_\alpha}$, with $n_1^i,  n_2^i \in \mathbb{Z}$.
Therefore, also the distance between two sites can be written as a linear combination of the primitive vectors. Namely
\begin{equation}\label{eq_distance}
{\rr_{ij{\alpha\beta}}} = {\x_{i{,\alpha}}} - {\x_{j{,\beta}}} = n_1^{ij} \ba_1 + n_2^{ij} \ba_2 {+{\bb_{\alpha\beta}} }, \qquad \text{with} \quad n_1^{ij}, n_2^{ij} \in \mathbb{Z} ,
\end{equation}
where $n_1^{ij} = n_1^i - n_1^j$, $n_2^{ij} = n_2^i - n_2^j$ { and ${\bb_{\alpha\beta}} 
 = \bb_\alpha -\mathbf{b_{\beta}}$}.
Then, the distance between two lattice sites can be written as a function of two integers {$n_1=n_1^{ij}$, $n_2=n_2^{ij}$ and a vector ${\cb_{\gamma}=\bb_{\alpha\beta}}$}. Namely, we have
\begin{equation}\label{eq.def.distancer}
    \rr(n_1, n_2{, \gamma})=n_1 \ba_1 + n_2 \ba_2 {+{\cb_{\gamma}}},
\end{equation}
with $n_1,n_2\in\mathbb Z$ and
\begin{equation}\label{eq-def.Db}
    \cb_{\gamma} \in \mathcal{D}_b = \{\, \bb_\alpha -{\bb_{\beta}} \mid \alpha,\beta = 1,\dots,K\}.
\end{equation}

Suppose that the lattice is finite, with $N_1$ and $N_2$  cells along the first and the second primitive vectors, respectively.
In this case, the coefficients for the position of the $i$-th site are $n_1^i \in [0, N_1-1]$ and $n_2^i \in [0, N_2-1]$.
Similarly, the coefficients that define the distance between the sites are $n_1^{ij} = (n_1^i - n_1^j) \in [-N_1+1, N_1-1]$ and $n_2^{ij} = (n_2^i - n_2^j) \in [-N_2+1, N_2-1]$.
As a result, the set of all possible distances between the lattice sites is
\begin{equation}\label{eq_distance_domain}
\mathcal{D} =\{ \rr(n_1, n_2{, \gamma})\mid n_i\in[-N_i+1,N_i-1], \cb_\gamma\in\mathcal{D}_b\},
\end{equation}
where the form of $\rr(n_1, n_2{, \gamma})$ is defined in Eq.~\eqref{eq.def.distancer}.

The weights $\omega(\rr)$ are given by the number of pairs of lattice cells connected by the vector $\rr$. The explicit value can be computed 
by accounting that the distance $\rr(n_1, n_2{, \gamma})$ links two cells that are separated by $n_i\ab_i$ along the $i$-th direction. Since the lattice is finite, there are $N_i-|n_i|$ pairs of cells at such separation. Therefore, we have
\begin{equation}\label{eq_weights}
    \omega(n_1, n_2,\gamma) = {\omega_{\gamma}}(N_1-|n_1|)(N_2-|n_2|),
\end{equation}
where $\omega_{\gamma}$ is the number of pairs of atoms in the same cell that are ${\cb_{\gamma}}$ apart. For example, for a lattice with a single atom per unit cell there is only one basis vector $\bb_1=(0,0)$ and only one possible (null) distance between atoms in the same cell. Thus, $\mathcal D_b=\{\cb_1=(0,0)\}$ and $\omega_1=1$. Conversely, for the graphene plate, which is the case of interest, one has $\bb_1=(0,0)$ and $\bb_2=a(\tfrac12,\tfrac{\sqrt{3}}{6})$. According to Eq.~\eqref{eq-def.Db}, we have
\begin{equation}
\begin{aligned}
        \mathcal D_b&=\{
   \cb_1=\bb_1-\bb_1,\cb_2=\bb_1-\bb_2,\cb_3=\bb_2-\bb_1,\cb_4=\bb_2-\bb_2 
    \},\\
    &=\{\cb_1=(0,0),\cb_2=-a(\tfrac12,\tfrac{\sqrt{3}}{6}),\cb_3=a(\tfrac12,\tfrac{\sqrt{3}}{6}),\cb_4=(0,0)
    \}.
\end{aligned}
\end{equation}
Since $\cb_4=\cb_1$, we can neglect $\cb_4$ and count twice $\cb_1$. Then, we find $\omega_1=2$ and $\omega_2=\omega_3=1$.

By considering the explicit form of the weights, Eq.~\eqref{eq_basic_fancy_sum} becomes
\begin{equation}
    \sum_{\rr \in \mathcal{D}} \omega(\rr) f(\rr, R_0, \mathbf{d}) =\sum_{n_1 = -N_1+1}^{N_1-1} \sum_{n_2 = -N_2+1}^{N_2-1} {\sum_{{\cb_{\gamma}} \in \mathcal{D}_b}} {\omega_{\gamma}}(N_1-|n_1|)(N_2-|n_2|) \, f(\rr(n_1, n_2{, \gamma}), R_0, {\db}) .
\end{equation}
Such an expression involves the sum of 
\begin{equation}\label{eq.linearscaling}
    (2 N_1 -1)  (2 N_2 -1) [1+K(K-1)] \simeq 4 N_1 N_2 [1+K(K-1)] = 4 \frac{[1+K(K-1)]}{K} N
\end{equation}
terms, where we used that the total number of atoms is given by the product of the number of cells per direction times the number of atoms per cell, i.e.~$N=N_1N_2K$.
These expressions are valid for any two-dimensional lattice, but can be easily generalized to higher dimensions.

As shown by Eq.~\eqref{eq.linearscaling}, the sum involves a number of terms that scales linearly with the number of atoms. Such a scaling needs to be compared with the quadratic one provided by Eq.~\eqref{eq_basic_sum}. Correspondingly, we obtain a speedup in the numerical summation for $N \gtrsim 20$, which reaches a factor of $10^{10}$ for $N\sim10^{10}$. Figure \ref{img_execution_time_vary_N} provides the comparison of the execution time for the numerical summation when employing Eq.~\eqref{eq.linearscaling} (blue dots) and Eq.~\eqref{eq_basic_sum} (red dots), {for $R_0=1\,$\AA\, and $d=10^6\,$\AA}. For the summation we employed {a simple in-house python code with \texttt{numba} just-in-time compiler~\cite{lam_numba_2015} on a personal laptop with Intel processor i7-8750H.} 
We extensively verified that the results of the two sums coincide to the same value, up to the machine precision for system with $N{ \leq 2 \times 10^5}$ {particles. Moreover, when we evaluated Eq.~\eqref{eq_basic_sum} for very large systems and $d \lesssim a $, we found that the accumulation of rounding errors produced an overall error that can reach the order of magnitude of the entire sum, thus making the estimation of the sum worthless.

\begin{figure}[h!] 
\centering
\includegraphics[width=0.65\textwidth]{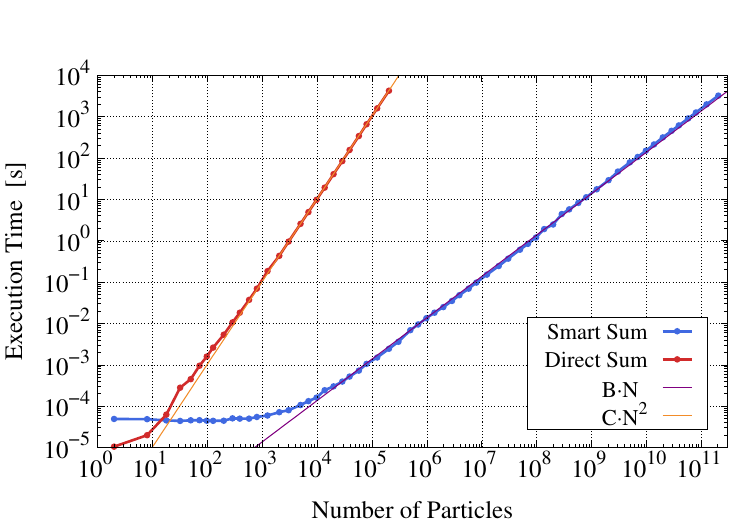} 
\caption{{Execution time for the numerical summation as a function of the number of particles $N$ for a  plate of graphene. Each point corresponds to a single evaluation Eq.~\eqref{eq.linearscaling} (blue dots and fitted line) and Eq.~\eqref{eq_basic_sum} (red dots and fitted line)}. The purple and orange lines provide  the scaling laws for Eq.~\eqref{eq.linearscaling} and Eq.~\eqref{eq_basic_sum} respectively.}
\label{img_execution_time_vary_N}
\end{figure}

\end{document}